\documentclass[12pt]{iopart}

\usepackage{multicol}
\usepackage{graphicx}
\usepackage{booktabs}
\usepackage{amssymb,bm,mathrsfs,bbm,amscd}
\usepackage{lastpage}
\usepackage{CJK}

\def\bd{\begin{document}} \def\ed{\end{document}}
\def\bmp{\begin{minipage}} \def\emp{\end{minipage}}
\def\bcc{\begin{center}} \def\ecc{\end{center}}     \def\npg{\newpage}
\def\beq{\begin{equation}} \def\eeq{\end{equation}} \def\hph{\hphantom}
 \def\r#1{$^{[#1]}$}
\def\n{\noindent} \def\ni{\noindent} \def\pa{\parindent}
\def\hs{\hskip} \def\vs{\vskip} \def\hf{\hfill} \def\ej{\vfill\eject}
\def\cl{\centerline} \def\ob{\obeylines}  \def\ls{\leftskip}
\def\underbar#1{$\setbox0=\hbox{#1} \dp0=1.5pt \mathsurround=0pt
   \underline{\box0}$}   \def\ub{\underbar}    \def\ul{\underline}
\def\f{\left} \def\g{\right} \def\e{{\rm e}} \def\o{\over} \def\d{{\rm d}}
\def\vf{\varphi} \def\pl{\partial} \def\cov{{\rm cov}} \def\ch{{\rm ch}}
\def\la{\langle} \def\ra{\rangle} \def\EE{e$^+$e$^-$} \def\pt{p_{\rm T}}
\def\pti{p_{{\rm T},i}} \def\yti{y_{{\rm T},i}}
\def\ptj{p_{{\rm T},j}}\def\mt{m_{\rm T}} \def\yt{y_{\rm T}} \def\vt{v_{\rm T}}

\def\bitz{\begin{itemize}} \def\eitz{\end{itemize}}
\def\btbl{\begin{tabular}} \def\etbl{\end{tabular}}
\def\btbb{\begin{tabbing}} \def\etbb{\end{tabbing}}
\def\beqar{\begin{eqnarray}} \def\eeqar{\end{eqnarray}}
\def\\{\hfill\break} \def\dit{\item{-}} \def\i{\item}
\def\bbb{\begin{thebibliography}{9} \itemsep=-1mm}
\def\ebb{\end{thebibliography}} \def\bb{\bibitem}
\def\bpic{\begin{picture}(260,240)} \def\epic{\end{picture}}
\def\akgt{\cl{\bf ACKNOWLEDGMENTS}}
\def\fgn{\noindent{\bf\large\bf figure captions}}

\def\lan{\langle}
\def\ran{\rangle}
\def\p{\pi}
\def\ifmath#1{\relax\ifmmode #1\else $#1$\fi}%
\def\rc{\ifmath{{\mathrm{c}}}}
\def\cut{\ifmath{{\mathrm{cut}}}}
\def\rF{\ifmath{{\mathrm{F}}}}
\def\rK{\ifmath{{\mathrm{K}}}}
\def\rp{\ifmath{{\mathrm{p}}}}
\def\rt{\ifmath{{\mathrm{t}}}}
\def\LAB{\ifmath{{\mathrm{LAB}}}}
\def\cut{\ifmath{{\mathrm{cut}}}}

\newcommand{\gguide}{{\it Preparing graphics for IOP journals}}
\begin{document}

\title[]{Influences of statistics and initial size fluctuation on high-order cumulants of conserved quantities in relativistic heavy ion collisions}

\author{Lizhu Chen$^1$, Zhiming Li $^2$ and Yuanfang Wu$^2$}
\address{$^1$ School of Physics and Optoelectronic Engineering, Nanjing University of Information Science and Technology, Nanjing 210044, China}
\address{$^2$ Key Laboratory of Quark and Lepton Physics (MOE) and Institute of Particle Physics, Central China Normal University, Wuhan 430079, China}
\pacs{25.75.Nq, 12.38.Mh, 21.65.Qr}

\begin{abstract}
By the generator of the UrQMD  model, event statistics for the products of kurtosis ($\kappa$) and variance ($\sigma^2$) of net-proton and net-charge multiplicity distributions are carefully studied. It is shown that the  statistics at RHIC/BES below $\sqrt {s_{NN}} < 19.6$ GeV are not sufficient for using the method of Centrality Bin Width Correction (CBWC). Corresponding results are systematically underestimated. A way to improve the CBWC method is proposed. It can remove the statistics dependence of the data and reduce the initial size fluctuation as well. 
\end{abstract}


\section{Introduction}
\vspace{1mm}

The ratios of high-order cumulants (from the 3rd to the 6th order ones) of conserved quantities are suggested as sensitive measurements of the QCD phase transition~\cite{diagram-1, diagram-2, diagram-3, star-1007, diagram-4, diagram-5, diagram-6, Nu-science}.  Recently, the ratios of low-order cumulants (from the 1st to the 3rd order ones) are turned out to be a useful probe of freeze-out temperatures~\cite{Karsch-frozen-out-1, Karsch-frozen-out-2}. So it is important to precisely determine the cumulants experimentally, and subtract the influence of non-critical effects. 

The initial size fluctuation is one of the important non-critical effects. Although it has not been taken into account in most of the theoretical calculations~\cite{fix-volume-1, fix-volume-2,fix-volume-3}, it exists in all event variables in experiments. Experimentally, the initial size is usually quantified by the collision centrality, which is determined by the multiplicity ($N_{ch}$) of the final state. A centrality bin corresponds to a range of multiplicity. For a given centrality bin, i.e., a range of multiplicity, the initial size still fluctuates from event to event~\cite{impact-1, impact-2}. Therefore,  its influence on high-order cumulants of conserved quantities cannot be neglected~\cite{henning-volume, skokov-volume}.  

In order to reduce the initial size fluctuation, it is suggested to calculate the cumulant at each of $N_{ch}$.
The cumulant is averaged over all multiplicities in a given centrality, where the average is weighted by the number of events in each of $N_{ch}$. This is  called the Centrality Bin Width Correction (CBWC)~\cite{cbwc-star}. The corrected cumulants  obviously show less centrality dependence~\cite{cbwc-star, xiaofeng-1302, cbwc-prc-nihar}.


A precise estimation of the high-order cumulant requires larger statistics. At RHIC/BES, however, the number of events, or statistics, is not yet large enough. In this case, we should carefully check if the statistics are large enough for the CBWC method.  

At the lowest RHIC/BES energy ($\sqrt{s_{NN}}=7.7$ GeV), the total available number of events for the analysis of net-proton and net-charge cumulants are only a few millions, and a few hundred millions at the top RHIC/BES energy ($\sqrt{s_{NN}}=200$ GeV)~\cite{C4-proton, star-charge}. The statistics in each of $N_{ch}$ are much less than the whole data sample. As a rough estimation, for 0-5\% centrality at $\sqrt{s_{NN}}=7.7$ GeV, the number of events for the analysis of net-proton is around 0.19 Million (M).~\footnote{The analyzed statistics for cumulants of net-proton multiplicity distributions is 3 Million at $\sqrt{s_{NN}}=7.7$ GeV in Au + Au collisions. Ignoring the efficiency loss at peripheral collisions, we just simply suppose that $\frac{1}{16}$ of the events are within the $0-5\%$ centrality. } There are around 100 multiplicity bins without considering the tail of the multiplicity distributions~\cite{prc-centrality-7.7}. Then the total number of events in the sample in each $N_{ch}$ is just around 1900.  Therefore, it should be carefully checked if these statistics are enough for the CBWC method.

As we have discussed, the motivation of the CBWC method is to reduce the initial size fluctuation, but calculating cumulants in each of $N_{ch}$ is not only the solution. In fact, many other factors, such as tracking inefficiency loss which can fluctuate event-by-event, and fluctuations in particle production, affect the multiplicity measurement. Those factors invalidate the one-to-one correspondence of $N_{ch}$ to initial size. If we select a proper multiplicity bin width which is approximately the size of the resolution of $N_{ch}$, those uncontrollable factors will be smeared. Since the width of this kind of multiplicity bin is slightly wider than each of multiplicity and much narrower than the entire multiplicity region in a given centrality, the statistics will be largely improved, and the initial size fluctuations will be reduced as well.

In this paper,  the application of the CBWC method for the statistics of RHIC/STAR experiments is carefully discussed by using the sample generated by the UrQMD model. It is found that the statistics at low RHIC/BES energies are not sufficient for applying the CBWC method, and the expectations of $\kappa\sigma^2$ of net-proton and net-electric charge are obviously smaller than those with sufficient statistics.

In Section III, way to improve the CBWC method is proposed. Instead of getting the cumulant in each of $N_{ch}$ in a given centrality, we show the statistics dependence of $\kappa\sigma^2$ at various centrality bin widths. An appropriate centrality bin width is found, where a stable mean of $\kappa\sigma^2$ is obtained and initial size fluctuation is well reduced with contemporary statistics at RHIC/STAR experiments. Finally,  the summary is provided in Section IV.  

\begin{figure}
\centering
\includegraphics[width=6.0in]{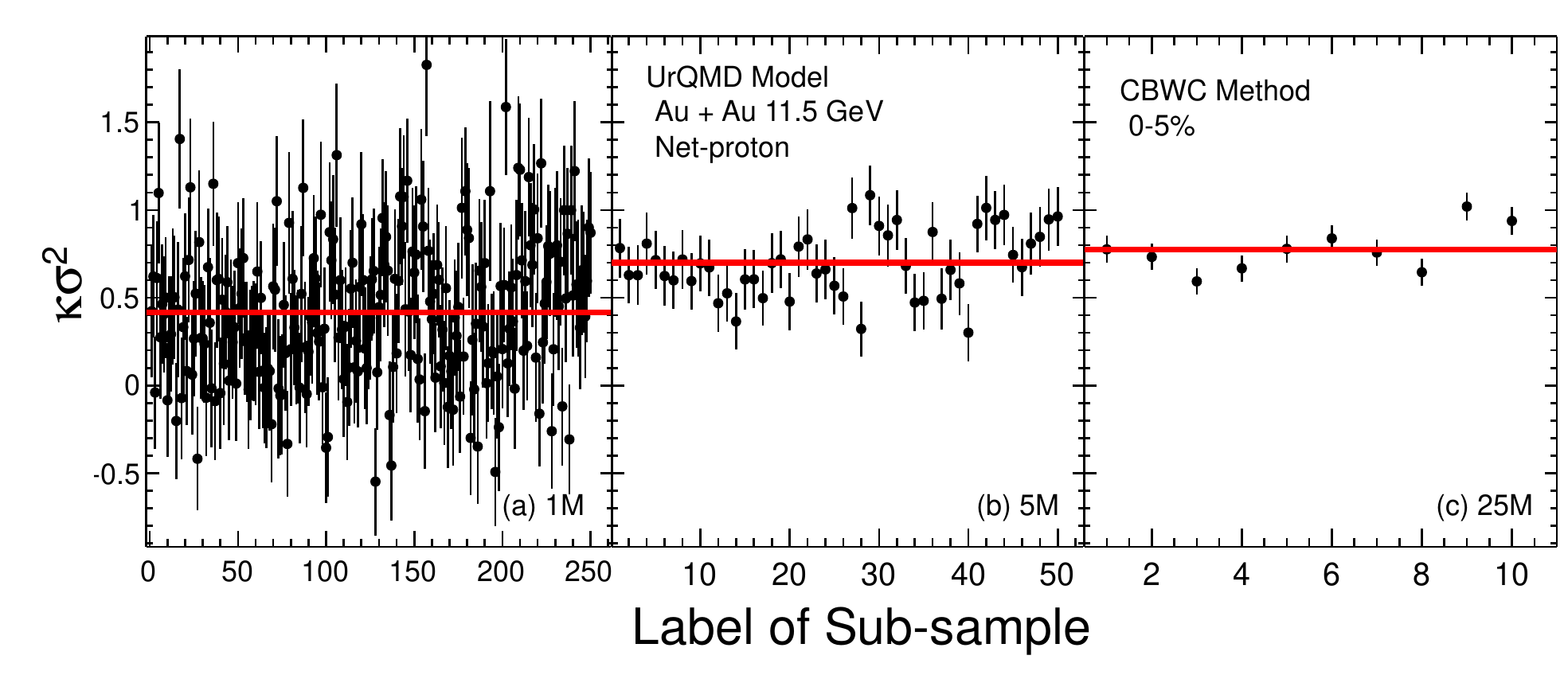}
\caption{\label{Fig-sample} (Color online) The fluctuations of $\kappa \sigma^2$ of net-proton calculated by the CBWC method for central collisions, with centrality $0-5\%$.  Where total numbers of minimum bias events for each  of sub-samples are 1M (a) , 5M (b) and 25M (c), respectively. The minimum bias sub-samples is Au + Au collisions at 11.5 GeV generated by UrQMD model. The x-axis is the label of sub-samples. The solid line in each panel is the mean of all points. } 
\end{figure}

\section{The problem of current calculations}

The sample generated by UrQMD model~\cite{UrQMD} is used to examine if the statistics of current RHIC/BES energies are enough for the CBWC method. Although the particle production mechanism of the UrQMD model may differ from that of a real experiment, statistics dependence of the cumulants should be a valuable reference for experimental data analysis.

We focus our discussions on the case of Au + Au collisions at $\sqrt{s_{NN}}= 11.5$ GeV. Its statistics are 6.6M for net-proton, and 2.4M for net-charge at RHIC/BES as showed in Table I ~\cite{C4-proton, star-charge}. We firstly simulate total 250 million minimum bias events, and then randomly divide the total events into 250, 125, 50, 25, 10 and 5 sub-samples. Corresponding statistics for each kind of sub-samples are 1M, 2M, 5M, 10M, 25M and 50M, respectively. So we can see how the mean value at each kind of sub-sample change with the statistics. These statistics cover all of the cases at RHIC/STAR shown in Table I.

\begin{table}[htmp]
\centering
\begin {tabular}{|c|c|c|c|c|c|c|c|} \hline
Energy (GeV)  & 200  & 62.4 & 39 & 27 & 19.6 & 11.5 & 7.7\\
\hline
Events of net-proton (Million)  & 238  & 47 & 86 & 30 & 15 & 6.6 & 3\\
\hline 
Events of net-charge (Million)  & 75  & 32 & 56 & 24 & 15.5 & 2.4 & 1.4\\
\hline
\end{tabular}
\caption{Statistics (0-80$\%$ centrality) at RHIC/BES energies for net-proton and net-charge cumulant analysis. The difference between net-proton and net-charge comes from event selections~\cite{C4-proton, star-charge}. \label{STAR-statistics}}
\end{table}

For a given sub-sample, we divide it into 9 centralities, the same way we did in data analysis~\cite{C4-proton, star-charge}. We choose the centrality 0-5\% as an example. Where, range of $N_{ch}$ for centrality definition is the largest, and the multiplicity distributions of net-proton (or net-charge) is the widest. So measured results are more sensitive to statistics than that of other centralities. 

According to the CBWC method, the $\kappa_i \sigma_i^2$ of net-proton multiplicity distribution is calculated in each of $N_{ch}$ and weighted by the number of events in correspondent $N_{ch}$. For any sub-sample with total 1M minimum bias events,  the results at each sub-sample are presented in Fig.~\ref{Fig-sample}(a) by black solid points.  The x-axis is the label of sub-samples. The red line is the mean value of $\kappa \sigma^2$ ($\left< \kappa \sigma^2\right>$) of net-proton, which is averaged over total 250 such kind of sub-samples. We can see that $\kappa_i \sigma_i^2$ is unstable and fluctuates around $\left< \kappa \sigma^2\right>$.
 
\begin{figure}[htmp]
\centering
\includegraphics[width=4.5in]{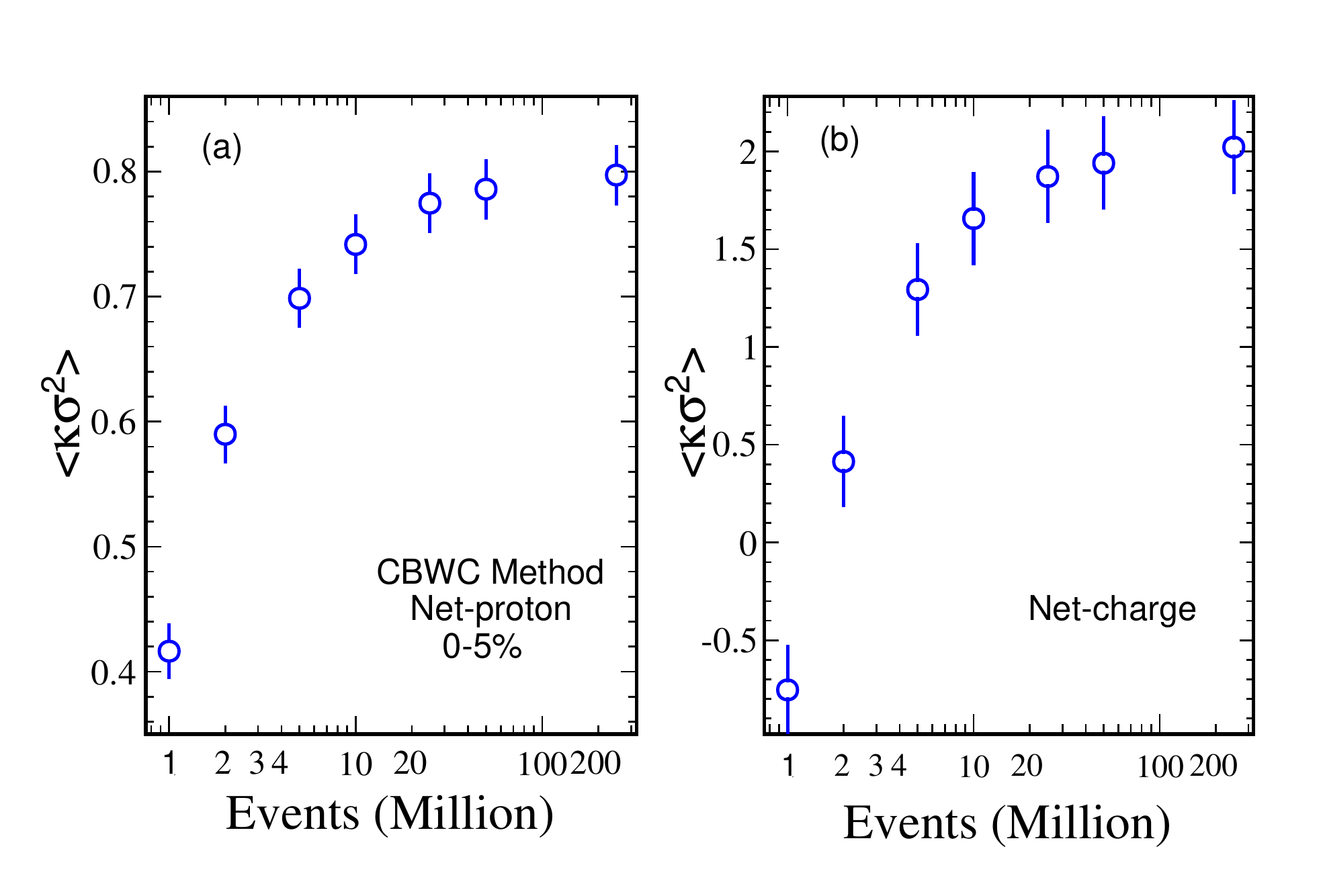}
\caption{\label{statistics-Nch} (Color online) Statistics dependence of $\left <\kappa\sigma^2\right>$ of net-proton and net-charge multiplicity distributions calculated by the CBWC method in centrality 0-5\%  at $\sqrt{s_{NN}}$=11.5 GeV by the UrQMD model. The x-axis is the number of the total minimum bias events of each sub-sample. The statistical errors are obtained from the formula of error propagation. The error of each point is very close to each other since the simulated data used for all the data points are identical except that they are chopped into different numbers of sub-samples.} 
\end{figure}

The results for the sub-samples with 5M and 25M minimum bias events are presented in Fig.~\ref{Fig-sample}(b) and (c), respectively.  Fig.~\ref{Fig-sample} shows $\left<\kappa \sigma^2\right>$ keeps increasing from Fig.~\ref{Fig-sample}(a) to (c), which indicates that 1M (or 5M) minimum bias events are not enough for using the CBWC method. 



In order to find the statistics required for a stable $\left<\kappa \sigma^2\right>$ of net-proton multiplicity distributions, the statistics dependence of $\left<\kappa \sigma^2\right>$ at centrality 0-5\% is shown in Fig.~\ref{statistics-Nch}(a). Where the x-axis is the number of total minimum bias events of each sub-sample.  
These errors are close to each other and correlated among the data points, because the simulation data used for all the data points are identical except that they are chopped into different numbers of sub-samples. Therefore, the differences between the data points are real despite they are smaller than the drawn error bars.
The results shown in this plot can be directly compared with the statistics at RHIC/STAR showed in Table I. The statistics we mentioned in the following are the minimum bias sample.  

Fig.~\ref{statistics-Nch}(a) shows that $\left<\kappa\sigma^2\right>$ increases rapidly with statistics when 
it is smaller than 10M. With further increases in statistics, the mean increases less and less and finally converges to a saturated value at 250M. If the difference to this saturated value less than 5$\%$ of it is acceptable (Considering that experimental uncertainties are often larger than 5\%, such as RHIC/STAR, 5$\%$ difference is reasonable.), the required statistics can be obtained from the plot. For net-proton, the differences in 10M and 25M are 7.0$\%$ and 2.8$\%$ to approach the saturation value, respectively. So the statistics with the difference less than 5$\%$ is about 15M. This statistics can be achieved when the incident energy is above 19.6 GeV at RHIC/BES, cf. Table I. The current statistics at low RHIC/BES energies are not enough for applying the CBWC method.  

Fig.~\ref{statistics-Nch}(a) also shows that the poorer the statistics, the smaller  $\left<\kappa \sigma^2\right>$ is. 
With insufficient statistics, $\left<\kappa \sigma^2\right>$ is systematically underestimated. This is more serious at lower incident energy region at RHIC/BES where the statistics is even poorer, such as a few millions events at 7.7 GeV. 

The statistics dependency of $\left<\kappa \sigma^2\right>$ for net-charge is presented in  Fig.~\ref{statistics-Nch}(b). It shows that more statistics is needed to approach the same standard as that for net-proton, i.e., the difference with the saturated value at 250M is less than $5\%$ of it. The differences in 25M and 50M are 7.4$\%$ and 4.1$\%$ of saturated value, respectively. The required statistics have to be about 50M, much more than that for net-proton multiplicity distributions. 

The statistics at RHIC/BES experiment are less than 50M below 27 GeV, cf., Table I. The lower the energy, the poorer the statistics is in experiment. The experimental statistics dependence of $\kappa \sigma^2$ of net-charge multiplicity distributions are not eliminated especially at 7.7 GeV shown in ~\cite{star-charge}, at which the central data point are obviously smaller than that at the other energies in centrality 0-5\%.

Here, we only show the statistics dependence of $\kappa \sigma^2$ of net-proton and net-charge multiplicity distributions. The required statistics should be observable dependent. In general, the higher the order of the cumulant, the larger the statistics is needed. It should be examined case by case carefully.



\section{A way to improve the CBWC method}

Fig.~\ref{statistics-Nch} shows that with increase in statistics, $\left<\kappa\sigma^2\right>$ becomes closer to the saturation value. The statistics for each of $\kappa_i \sigma_i^2$ will be largely increased if we do not calculate $\kappa_i \sigma_i^2$ in each of $N_{ch}$, but a properly wider multiplicity bin, which is wider than each of $N_{ch}$ and smaller than the whole multiplicity range.  In this case, if the initial size fluctuation can still be reduced equally as that calculated in each of $N_{ch}$, it will be a better solution for
applying the CBWC method to current RHIC/BES with the limited statistics resource. 

\begin{figure} [htmp]
\centering
\includegraphics[width=4.6in]{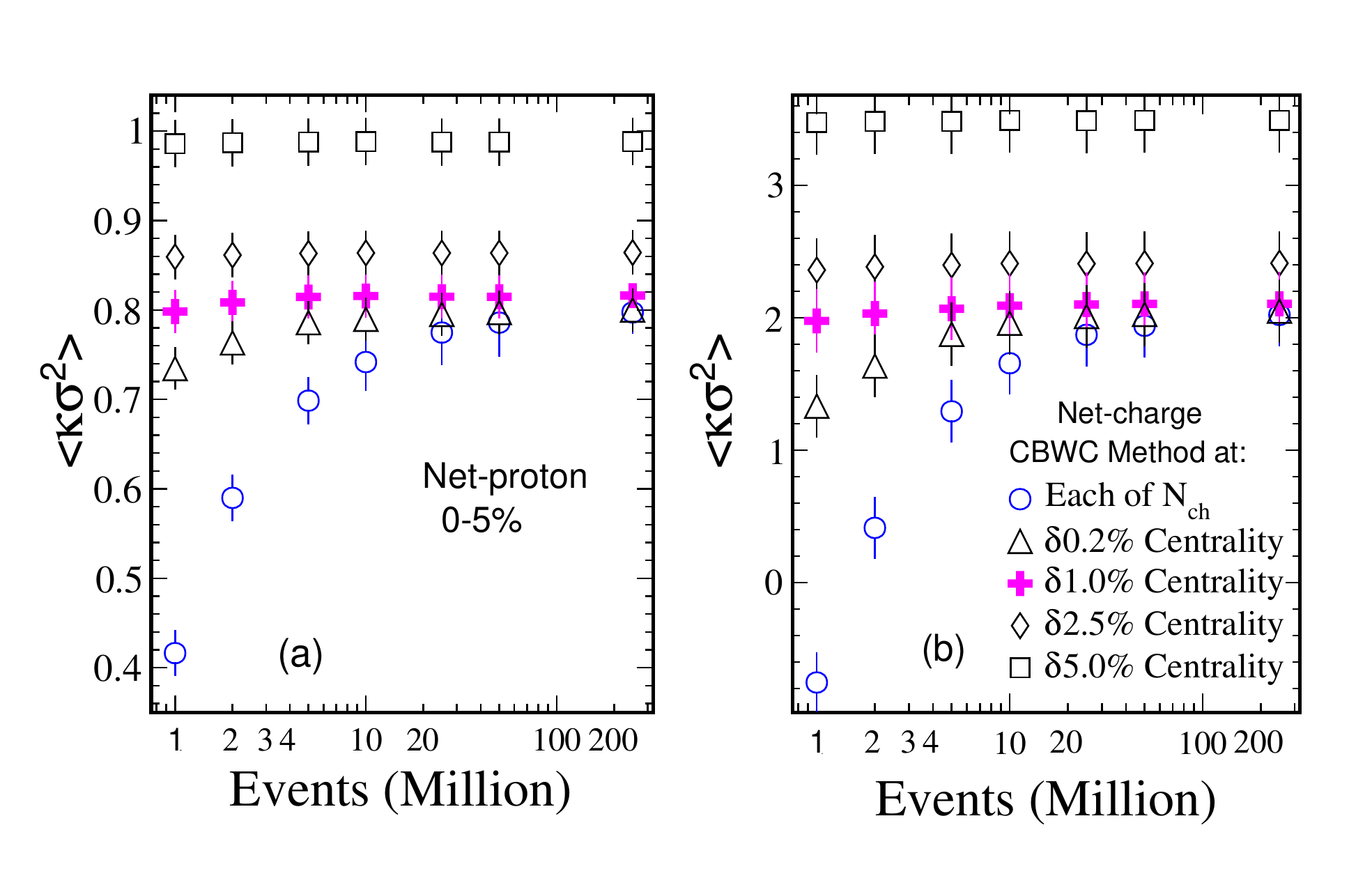} 
\caption{ \label{statistics-width}  (Color online) Statistics dependence of means of $\kappa \sigma^2$ of net-proton (a) and net-charge (b) multiplicity distributions calculated by the CBWC method at various centrality bin widths: $\delta5.0\%$ (black solid triangles),  $\delta2.5\%$ (black open squares),  $\delta$1.0\% (pink solid crosses), $\delta$0.2\% (black open triangles) and each of $N_{ch}$ (blue open circles)), in centrality $0-5\%$. Where the minimum bias sub-sample is Au + Au collisions at $\sqrt{s_{NN}}=$11.5 GeV generated by the UrQMD model.} 
\end{figure}

In order to study if such a proper multiplicity bin exists for current statistics at RHIC/BES, we divide the whole multiplicity range in centrality $0-5\%$ into 25 bins ($\delta$0.2\% centrality bin width), 5 bins ($\delta$1.0\% centrality bin width), 2 bins ($\delta$2.5\% centrality bin width) and 1 bin ( $\delta$5.0\% centrality bin width), respectively. The larger dividing bin number, the smaller statistics in each bin. By previously generated 250M minimum bias sample, we still randomly divide them into 250, 125, 50, 25, 10 and 5 sub-samples, respectively. Fig.~\ref{statistics-width} presents the statistics dependence  of $\left< \kappa \sigma^2 \right>$ with various different proposed centrality bin width: each of $N_{ch}$ (blue open circles), $\delta$0.2\% (black open triangle), $\delta$1.0\% (pink solid cross), $\delta$2.5\% (black open diamond), and $\delta$5.0\% (black open square), respectively.

It shows that with the increase of centrality bin width, $\left<\kappa \sigma^2 \right>$ becomes  flatter, i.e., the mean become independent of the statistics. For centrality bin width $\delta1.0\%$, as showed by the pink solid crosses in the figure, the current statistics at RHIC/BES, a few millions, are enough for getting a reliable result. Meanwhile, its saturated mean is still close to that calculated at each of multiplicity with sufficient statistics, i.e., the pink solid crosses are closer to the blue open circles when statistics is above 15M in Fig.~\ref{statistics-width}(a) and 50M in Fig.~\ref{statistics-width}(b). This means that the initial size fluctuations are equally reduced. With further increase of centrality bin width, e.g., $\delta2.5\%$, or $\delta5.0\%$, the saturated mean is obviously higher than that obtained from $\delta1.0\%$ centrality bin width. This shows that the initial size fluctuations are not completely eliminated. So a proper centrality bin width is about $\delta 1.0\%$. 

\begin{figure}[htmp]
\centering
\includegraphics[width=2.8in]{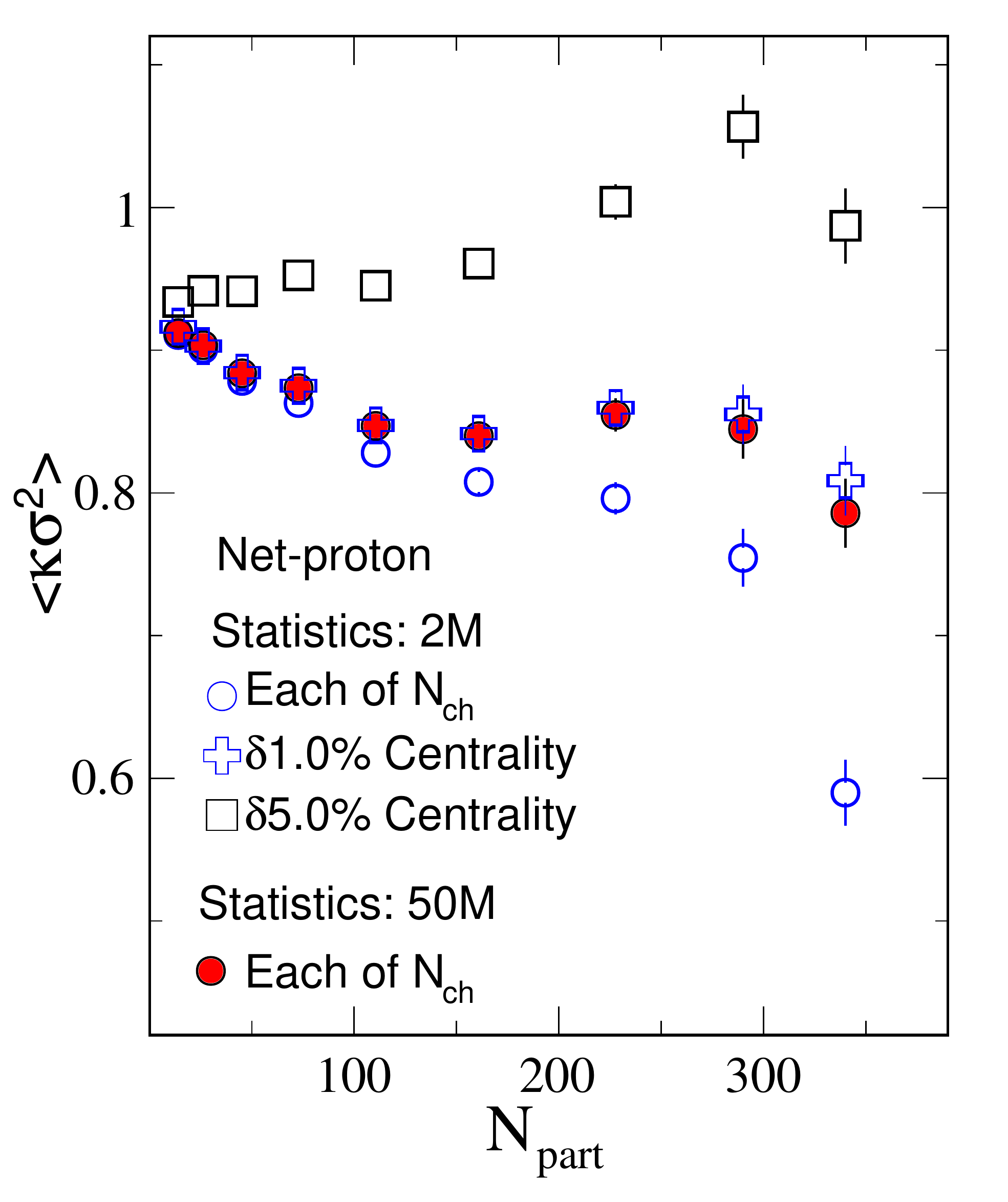}
\caption{\label{statistics-centrality} (Color online) Centrality dependence of $\left< \kappa \sigma^2 \right> $ of net-proton multiplicity distributions calculated by the CBWC method with different centrality bin widths: $\delta 5.0\%$ (black open square),  $\delta$1.0\% (blue open cross), and each of $N_{ch}$ (blue open circles), for total number of events 2M. The red solid circles are calculated by the CBWC method at each of $N_{ch}$ for the total number of events 50M. The sub-samples are Au + Au collisions at 11.5 GeV produced by the UrQMD model.} 
\end{figure}

To check if this centrality bin width works for other centralities, the centrality dependence of $\left<\kappa \sigma^2\right>$ of net-proton calculated by the CBWC method for different centrality bin widths, $\delta5.0\%$ (black open square),  $\delta1.0\%$ (blue open cross), and each of $N_{ch}$ (blue open circle), with total number of events 2M are presented in Fig.~\ref{statistics-centrality}. The results calculated by the CBWC method at each of $N_{ch}$ with sufficient statistics, (i.e., the total number of events 50M) are also presented by the red solid circles. They can be viewed as a standard of comparison.


Fig.~\ref{statistics-centrality} shows $\left<\kappa \sigma^2\right>$ calculated at $\delta$1.0\% centrality bin width with 2M events, which is the same magnitude as current statistics at RHIC/BES, are almost overlapped with those red solid circles. This indicates that $\delta$1.0\% centrality bin width is proper for nine centralities. 

Fig.~\ref{statistics-centrality} also shows that in central and mid-central collisions, the $\left <\kappa \sigma^2\right>$ calculated at each of $N_{ch}$ with 2M events are obviously smaller than those with sufficient statistics. While they almost coincide in peripheral collisions. These deviations and coincidences imply that the statistics of current RHIC/BES are not enough for applying the CBWC method in mid-central and central collisions, but enough in peripheral collisions. Meanwhile, we can see that the results with centrality bin width $\delta$5.0\%  are systematically larger than those with $\delta$1.0\%. It indicates that the initial size fluctuations are poorly reduced if the centrality bin width is too larger.

It should be noticed that the centrality dependence of $\left <\kappa \sigma^2\right>$ calculated by $\delta1.0\%$ centrality bin width in Fig.~\ref{statistics-centrality} is not completely flat. 
It may be a result of the poor centrality resolution in peripheral collisions as discussed in~\cite{xiaofeng-1302}. Moreover, the centrality can be defined by the number of participant nucleons, or the impact parameter, or the reference multiplicity. Each of them fluctuates from one to another. So the initial size fluctuation is hard to be eliminated completely, even if the CBWC method is used. It needs further investigation. We should be very careful in discussing the physics behind the currently measured $\kappa \sigma^2$.

Here, we only show the appropriate centrality bin width for $\kappa \sigma^2$ of net-proton and net-charge multiplicity distributions. It may be different from other orders of cumulants. 

\section{Summary}

In this paper, the influence of statistics and initial size fluctuation on the measurement of $\kappa \sigma^2$ is demonstrated. By the samples generated by the UrQMD model, statistics dependence of $\kappa \sigma^2$ of net-proton and net-charge multiplicity distributions at RHIC/BES energies is presented accordingly. It is shown that the statistics at RHIC/BES below $\sqrt {s_{NN}} < 19.6$ GeV are not sufficient for using the CBWC method. Corresponding results are systematically underestimated.  

An improved CBWC method is proposed. Where the measurement for $\kappa \sigma^2$ at each of $N_{ch}$ is substituted by that at a proper centrality bin width. We found that such a proper bin width is around $\delta 1.0\%$ for $\kappa \sigma^2$ of net-proton and net-charge multiplicity distributions. It can remove the statistics dependence of the data and reduce the initial size fluctuation as well. For the statistics of current RHIC/BES energies, such obtained $\left<\kappa \sigma^2\right>$ are as good as those estimated by the CBWC method with sufficient statistics. So it provides a more reliable measurement of $\left<\kappa \sigma^2\right>$.

\section{Acknowledgements}

This work is supported in part by the Scientific Research Startup Foundation of Nanjing University of Information Science and Technology (Grant No. 2013x025),  the Major State Basic Research Development Program of China under Grant No. 2014CB845402, the NSFC of China under Grants  No. 11221504, No. 11005046, and the Ministry of Education of China with Project No. 20120144110001.

\clearpage

\end{document}